\documentclass[prc,twocolumn,showpacs,preprintnumbers,amsmath,amssymb]{revtex4}

\usepackage{graphicx}

\usepackage{dcolumn}
\usepackage{bm}

\begin{document}


\title{Low-momentum nucleon-nucleon 
interaction and its application to the 
few-nucleon systems}


\author{S. Fujii}
\email{sfujii@nt.phys.s.u-tokyo.ac.jp}
\affiliation{%
Department of Physics, University of Tokyo, Tokyo 113-0033, Japan
}%
\author{E. Epelbaum}
\email{epelbaum@jlab.org}
\affiliation{Jefferson Laboratory, Newport News, VA 23606, USA}
\author{H. Kamada}
\email{kamada@mns.kyutech.ac.jp}
\author{R. Okamoto}
\email{okamoto@mns.kyutech.ac.jp}
\author{K. Suzuki}
\email{suzuki@mns.kyutech.ac.jp}
\affiliation{
Department of Physics, Kyushu Institute of Technology,
Kitakyushu 804-8550, Japan
}
\author{W. Gl\"ockle}
\email{Walter.Gloeckle@tp2.ruhr-uni-bochum.de}
\affiliation{
Institut f\"ur Theoretische Physik II, Ruhr-Universit\"at Bochum, 
D-44780 Bochum, Germany
}

\date{\today}

\begin{abstract}

Low-momentum nucleon-nucleon interactions are derived within the framework of
a unitary-transformation theory,
starting with realistic nucleon-nucleon interactions.
A cutoff momentum $\Lambda$ is introduced to specify a border between
the low- and high-momentum spaces.
By the Faddeev-Yakubovsky calculations the low-momentum interactions are
investigated with respect to the dependence of ground-state energies of
$^{3}$H and $^{4}$He on the parameter $\Lambda$.
It is found that we need the momentum cutoff parameter
$\Lambda \geq 5$ fm$^{-1}$ in order to reproduce satisfactorily the exact
values of the binding energies for $^{3}$H and $^{4}$He.
The calculation with $\Lambda =2$ fm$^{-1}$ recommended by Bogner {\it et al.}
leads to considerable overbinding at least for the few-nucleon systems.

\end{abstract}


\maketitle

\section{\label{sec:introduction}Introduction}

One of the fundamental problems in nuclear structure calculations is to
describe nuclear properties, starting with realistic nucleon-nucleon (NN)
interactions.
However, since this kind of interaction has a repulsive core
at a short distance,
one has to derive an effective interaction
in a model space from the realistic interaction,
except for the case of precise few-nucleon structure calculations.

Recently, Bogner {\it et al.} have proposed a low-momentum
nucleon-nucleon (LMNN) interaction which is constructed in momentum space
for the two-nucleon system from a realistic nucleon-nucleon interaction,
using conventional effective interaction techniques or renormalization group
ones~\cite{Bogner03}.
In the construction of the LMNN interaction a cutoff momentum $\Lambda$ is
introduced to specify a border between the low- and high-momentum spaces.

The LMNN interaction is constructed in order to account for the short-range
correlations of the two nucleons interacting in the vacuum.
So the question is to what
extent the obtained LMNN interaction is a good approximation also for
describing correlation of nucleons interacting in a nuclear many-body medium. 
The medium effect could appear through the single-particle potential
and three-or-more-body correlations.
From a practical point of view it is of high interest to explore
the sensitivity of calculated results to the cutoff momentum $\Lambda$.

Bogner {\it et al.} have constructed their LMNN interaction in a way to
conserve in the low-momentum region not only the on-shell properties
of the original interaction (i.e. phase shifts and the deuteron binding energy)
but also the half-on-shell $T$ matrix~\cite{Bogner03}.
They found that the LMNN interactions for $\Lambda=2.1$ fm$^{-1}$
corresponding to $E_{\rm lab}\simeq 350$ MeV become nearly universal, not
(or only weakly) dependent on the choice of realistic interactions employed.
They suggested to use their LMNN interaction directly in nuclear structure
calculations, such as the shell-model~\cite{Bogner02} and the Hartree-Fock
calculations~\cite{Coraggio03}.
They  claimed that for the momentum cutoff in the vicinity of
$\Lambda =2.1$ fm$^{-1}$ the calculated low-lying spectra for $^{18}$O,
$^{134}$Te, and $^{135}$I  are in good agreement
with the experimental data and depend weakly on $\Lambda$.
Their results have been found to agree with the data as good 
or even slightly better than
the results based upon the $G$ matrix which is constructed by taking into
account the short-range correlations, the Pauli blocking effect,
and the state dependence for each nucleus.

Kuckei {\it et al.}~\cite{Kuckei} investigated the nuclear matter
and the closed-shell nucleus $^{16}$O by using the LMNN interaction
obtained by Bogner {\it et al.}
They concluded that the LMNN interaction can be a very useful tool for
low energy nuclear structure calculations, 
and
one should be cautious
if the observable of interest is sensitive to the single-particle spectrum
at energies above the cutoff momentum.

However, one had better compare with the exact solutions once for all.
In cases of three- and
four-nucleon systems we can directly make such a comparison by solving the
Faddeev-Yakubovsky  equations~\cite{Faddeev}.

We investigate the LMNN interaction by means of unitary transformation
using two independent (but equivalent) approaches.
One approach is based on the unitary transformation of
the \=Okubo form~\cite{Okubo54}
in which a LMNN interaction is obtained from the scattering amplitude
in momentum space~\cite{Epelbaoum98,Epelbaoum99,Gloeckle82}.
Another  is  the unitary-model-operator approach
(UMOA)~\cite{Suzuki94,Fujii04}.
In contrast to the  $G$-matrix theory the UMOA 
leads to an energy-independent and Hermitian
effective interaction in a many-body system.
Contrary to the LMNN of Ref.~\cite{Bogner03}, we will not require
conservation of the half-on-shell $T$ matrix, which does not 
represent an observable
quantity. Only low-momentum NN observables such as the on-shell $T$
matrix, phase shifts and binding energies are guaranteed to remain 
unchanged under an  unitary transformation. 
In principle, one could achieve the equivalence of the half-on-shell
$T$ matrix by performing an additional unitary transformation in the
low--momentum space. We, however, refrain from doing that 
since we do not see any conceptual or practical advantage in requiring
the equivalence of the  half-on-shell $T$ matrix. 
 
In the present study we first apply the above-mentioned methods to the
two-nucleon system in momentum space to 
construct the LMNN interaction.
Although both approaches are based on the same idea of an  unitary
transformation, the calculation procedures for deriving the LMNN
interaction are independent and quite different from each other.
We calculate selected properties of the two-nucleon system using both
schemes to confirm the numerical accuracy. 
This cross check is useful to ensure the reliability of the 
numerical calculations. Secondly,  we investigate the $\Lambda$
dependence in structure calculations of few-nucleon systems, where
(numerically) exact calculations can be
performed~\cite{Kamada01,nogga03a}, and discuss the validity 
of the LMNN interaction by comparing the obtained results with the 
$exact$ values.

This paper is organized as follows.
In Sec.~\ref{sec:transformation}, after the basic formulation of the unitary
transformation is presented,
the two methods are given with emphasis on the different calculation procedures
for deriving the LMNN interaction.
In Sec.~\ref{sec:results}, LMNN interactions are constructed using both methods
from realistic nucleon-nucleon interactions such as
the CD-Bonn~\cite{Machleidt96} and the Nijm-I~\cite{Stoks94} potentials.
Then, the Faddeev and Yakubovsky equations are solved using
the LMNN interactions for various values of the cutoff parameter $\Lambda$.
Finally, we summarize our results in Sec.~\ref{sec:summary}.

\section{\label{sec:transformation}Unitary transformation of the Hamiltonian
for the two-nucleon system in momentum space}

We consider a quantum mechanical
system  described by a Hamiltonian $H$. The Schr\"odinger equation reads
\begin{equation}
\label{eq:scheq-1}
H\Psi =E \Psi .
\end{equation}
Introducing a unitary transformation $U$ with $UU^{\dagger}=1$
we obtain a transformed Schr\"odinger equation
\begin{equation}
\label{eq:transformed-1}
H'\Psi' =E\Psi'
\end{equation}
with the transformed Hamiltonian and state,
$H' = U^{\dagger}HU$ and  $\Psi'=U^{\dagger}\Psi$, respectively.
We introduce the concept of the effective Hamiltonian by means of
the partition technique.
The Hilbert space is divided into two subspaces,
the model space (P space) and its complement (Q space),
such that the Schr\"odinger equation
becomes a $2 \times 2$ block matrix equation
\begin{equation}
\label{eq:blocked-1}
\left[
 \begin{array}{cc}
  PH'P & PH'Q \\
  QH'P & QH'Q
 \end{array}
\right] 
\left[
 \begin{array}{c}
  P\Psi'\\
  Q\Psi'
 \end{array}
\right] 
=E
\left[
 \begin{array}{c}
  P\Psi'\\
  Q\Psi'
 \end{array}
\right] .             
\end{equation}
Here $P$ and $Q$ are the projection operators of a state onto the model space
and its complement, respectively, and they satisfy
$P+Q=1$, $P^{2}=P$, $Q^{2}=Q$, and $PQ=QP=0$.
The Q-space state is easily eliminated to produce the projected Schr\"odinger
equation
\begin{eqnarray}
\label{eq:projectedsch}
\left[\displaystyle
\begin{array}{c}
{ 
PH'P + PH'Q {1\over  E - QH'Q}
QH'P}
\end{array}
\right]
(P\Psi ') = E(P\Psi').
\end{eqnarray}
In general, the effective Hamiltonian which is given in parentheses
depends on the energy $E$ to be determined.
However, if the decoupling equation 
\begin{equation}
\label{eq:decouplingeq}
QH'P=0
\end{equation}
is satisfied,
then we have the equation for the energy independent effective Hamiltonian
in the P space
\begin{equation}
PH'P(P\Psi') = E(P\Psi').
\end{equation}
A unitary transformation can be parametrized as 
\begin{equation}
\label{eq:Okubo}
U = \left( \begin{array}{cc} P (1 + \omega ^\dagger \omega  )^{- 1/2}
P & - P
\omega ^\dagger ( 1 + \omega \omega ^\dagger )^{- 1/2}  Q \\
Q \omega  ( 1 + \omega ^\dagger \omega  )^{- 1/2} P
& Q (1 +  \omega \omega ^\dagger )^{- 1/2} Q
\end{array} \right),
\end{equation}
and the wave operator $\omega$ satisfies the condition
$\omega  = Q \omega  P$.
Equation~(\ref{eq:Okubo}) is well-known as the \=Okubo form~\cite{Okubo54}.
Notice that the unitary transformation given in Eq.~(\ref{eq:Okubo})
is by no means unique: in fact one can construct infinitely many different 
unitary transformations
which decouple the P and Q subspaces. For example, performing subsequently any
additional transformation, which is unitary
in the  P subspace, one would get a different LMNN interaction~\cite{Holt04}. 
The transformation in Eq.~(\ref{eq:Okubo})
depends only on the operator $\omega$ which mixes the P and Q
subspaces and is in some sense ``the minimal possible'' 
unitary transformation. For more discussion the reader is referred to
Ref.~\cite{Okubo54}.

In the present study, the above unitary transformation is used
in two different methods to derive LMNN interactions.
In the following sections, we shall give details on the two methods.

\subsection{\label{sec:method1}Method-1}

Consider a momentum-space Hamiltonian for the two-nucleon system of the form
\begin{equation}
H (\vec{p},\vec{p}\, '  ) =  H_0 (\vec{p}\, ) \, 
\delta (\vec{p} - \vec{p}\,'  ) 
+  V  (\vec{p},\vec{p}\, ' ) \, ,
\end{equation} 
where ${H}_0  (\vec{p}\, )= \vec{p} \, ^2/(2 M)$
with the reduced mass $M$ stands for the kinetic energy,
and $V  (\vec{p},\vec{p}\, ' )$ is the bare two-body interaction.
Our aim is to decouple the low- and high-momentum components of this
two-nucleon potential using the method of unitary transformation.
To achieve that, we introduce the projection operators
\begin{eqnarray}
P &=& \int d^3 p \,
| \vec{p} \,\rangle \langle {\vec p}\, | \,\, , \quad 
| \vec{p}\,|  \le  \Lambda \,\, , \nonumber \\
Q    &=& \int d^3 q \, 
| \vec{q} \,\rangle \langle {\vec q}\, | \,\, , \quad 
| \vec{q}\,| > \Lambda \,\, ,   
\label{eq:PQ}
\end{eqnarray}
where $\Lambda$ is a momentum cutoff whose value will be specified later,
and $P$ ($Q$) is a projection operator onto low- (high-) momentum states.
Based on the unitary-transformation operator given in Eq.~(\ref{eq:Okubo}),
the effective Hamiltonian in the P space takes the form
\begin{eqnarray}
\label{14}
P  H ' P  &=& P  ( 1 + \omega ^\dagger \omega )^{-1/2}
\left( H + \omega ^\dagger H + H \omega  + \omega ^\dagger H \omega 
\right) \nonumber \\
&\times &  (1 + \omega ^\dagger \omega )^{-1/2} P .
\end{eqnarray}
This interaction is by its very construction Hermitian~\cite{Okubo54}.

The requirement of decoupling the two spaces leads to the following nonlinear
integral equation for the operator $\omega$
\begin{eqnarray}
{ V} (\vec{q} ,\vec{p}\, ) &-& \int  d^3p' \, 
\omega ( \vec{q}, \vec{p}\, ' ) { V}( \vec{p}\,' , \vec{p}\,) 
\nonumber\\ &+&\int  d^3q' \, 
{ V} (\vec{q} , \vec{q}\,' ) \omega  (\vec{q}\,' , \vec{p}\, )
  \nonumber\\
&-& \int  d^3p' \,  d^3q' \, 
\omega (\vec{q} , \vec{p}\,' ) { V} (\vec{p}\, ' , \vec{q}\, ' ) 
\omega (\vec{q}\, ' , \vec{p}\, )
\nonumber\\
&=& (E_{{p}} - E_{{q}}) \, \omega (\vec{q} , \vec{p} \,)  ,
\end{eqnarray}
where we have denoted by $\vec{p}$ $(\vec{q})$ a momentum of the
P space (Q space) state and by $E_p$ ($E_q$) the kinetic energy $E_p
=p^2/(2 M)$ ($E_q =q^2/(2 M)$).

Alternatively, one can determine the operator $\omega$ from the following
linear equation, as exhibited in Refs.~\cite{Epelbaoum98,Epelbaoum99}:
\begin{equation}
\label{linear}
\omega  (\vec{q}, \vec{p} ) = \frac{T (\vec{q}, \vec{p}, E_p)}{E_p - E_q}
- \int  d^3 p '  \, \frac{ \omega  ( \vec{q}, \vec{p\,}' )
\; T (\vec{p\,}', \vec{p}, E_p )}
{E_p - E_{p '} + i \epsilon }.
\end{equation}
Here the integration over $p '$ goes from 0 to $\Lambda$.
Consequently, the dynamical input in this method is the $T$ matrix $T
(\vec{p_1}, \vec{p_2}, E_{p_2} )$.
Note that this is not a usual equation of the Lippmann-Schwinger (LS) type
since the position of the pole $E_p$ in the integration over $p '$ is not
fixed but moves with $p$.
In solving the integral equation Eq.~(\ref{linear}),
the second argument $p$ in $\omega$ varies, whereas the first one $q$
is a parameter.

In this study we have used the linearised equation Eq.~(\ref{linear}) to
project out high-momentum components from the realistic potentials.
Since the projection operators $P$ and $Q$ do not carry any angular dependence,
the integral equation Eq.~(\ref{linear}) can be solved for each partial wave
independently.
In the partial wave decomposed form it reads
\begin{eqnarray}
\label{linear_swave}
\omega ^{sj}_{ll'} (q, p) &=& \frac{T^{sj}_{ll'} (q, p, E_q)}{E_p - E_q} \\
&-& \sum_{\tilde l} \int_0^\Lambda p^{'2} \, d p '  \, \frac{ \omega ^{sj}_{l
\tilde l} (q, p' )
\; T^{sj}_{\tilde l l'} (p', p, E_p )}
{E_p - E_{p '} + i \epsilon}  ,\nonumber 
\end{eqnarray}
where  ${V}^{sj}_{ll'} ({q} ,{p}\, ) \equiv \langle lsj,  q | V | l'sj,
p\rangle$ and 
${\omega }^{sj}_{ll'} ({q} ,{p}\, ) \equiv \langle lsj,  q | \omega  | l'sj,
p\rangle$.
In the uncoupled case $l$ is conserved and equals $j$. In the coupled
cases it takes the values $l=j \pm 1$.

Equation~(\ref{linear}) and, consequently,
also Eq.~(\ref{linear_swave}) have a so-called moving singularity,
which makes it more difficult to handle than the LS equation.
Indeed, one has to discretize both $p$ and $p'$ points in
Eq.~(\ref{linear_swave}).
This does not necessarily allow to solve Eq.~(\ref{linear_swave}).
The difference $E_p-E_{p'}$ can be exactly zero since $p$ and $p'$
belong now to the same set of quadrature points.
Thus, one cannot calculate the principal value integral
in the same manner as for the LS equation.
To solve this equation we have used a method proposed by Gl\"ockle
{\it et al}.~\cite{Gloeckle82}.

Another problem arising in solving the linear equation is caused by the
fact that the driving term $T^{sj}_{ll'} (q, p, E_p)/(E_p - E_q)$ becomes very
large when $p \equiv | \vec p \,|$ and $q \equiv | \vec q \,|$ go to $\Lambda$.
Consequently, the equation becomes ill-defined.
To handle this problem we have regularized this equation by multiplying
the original potential $V (\vec{k} \, ',\vec k \,)$ with some smooth functions
$f(k ')$ and $f(k)$ which are zero in a narrow neighborhood of the points
$k '= \Lambda$ and $k = \Lambda$~\cite{Epelbaum_thesis}.
The precise form of this regularization does, in fact,
not matter~\cite{Epelbaoum99}.

Having determined the operator $\omega$, one can calculate the effective
Hamiltonian in the P space according to Eq.~(\ref{14}).
To evaluate the operator $P ( 1 + \omega ^\dagger \omega )^{-1/2}$
entering this equation we first diagonalize the operator
$(1 + \omega ^\dagger \omega )$ and then take the square root.
Finally, the effective potential can be found by subtracting the
(not transformed) kinetic energy term from the effective Hamiltonian.

\subsection{\label{sec:method2}Method-2}

In Eq.~(\ref{eq:Okubo}), the unitary-transformation operator $U$ has been given
by the block form with respect to the projection operators $P$ and $Q$.
We notice here that the operator $U$ can also be written in a more compact
form as~\cite{Suzuki82}
\begin{equation}
\label{eq:U}
U=(1+\omega-\omega ^{\dagger})
(1+\omega \omega ^{\dagger} +\omega ^{\dagger}\omega )^{-1/2}.
\end{equation}
Using the above operator $U$, in general,
the effective interaction $\tilde{V}$ for a many-nucleon system is
defined through
\begin{equation}
\label{eq:V_eff}
\tilde{V}=U^{-1}(H_{0}+V)U-H_{0},
\end{equation}
where $H_{0}$ is the kinetic energy of the constituent nucleons in the nuclear
system, and $V$ is the bare two-body interaction between the nucleons.
Here we apply thus defined effective interaction to the two-nucleon problem.
In that case $H_{0}$ becomes the relative kinetic energy of the two nucleons,
and $V$ the bare two-body interaction between the two nucleons.
Then the LMNN interaction $V_{{\rm low}-k}$ of interest in the present work is
given by
\begin{equation}
\label{eq:V_lowk}
V_{{\rm low}-k}= P_{{\rm low}-k}\tilde{V}P_{{\rm low}-k},
\end{equation}
where $P_{{\rm low}-k}$ is the projection operator
onto the low-momentum space for relative two-body states,
and is the same as $P$ in Eq.~(\ref{eq:PQ}) of the method-1.
In order to obtain $V_{{\rm low}-k}$ in the form of the matrix elements
using the plane-wave basis states, we shall present in the following a
procedure for the numerical solution.

We first consider an eigenvalue equation for the relative motion of a
two-nucleon system as
\begin{equation}
\label{eq:eigenvalue_eq}
(H_{0}+V)|\Psi _{n}\rangle =E_{n}|\Psi _{n}\rangle .
\end{equation}
The above equation is written also in an integral form concerning
relative momenta $k$ and $k'$ as
\begin{equation}
\label{eq:eigenvalue_eq_k}
\int ^\infty _0 \langle k'|H_{0}+V|k\rangle \langle k|
\Psi _{n}\rangle k^{2}dk=E_{n}\langle k'|\Psi _{n}\rangle,
\end{equation}
where
\begin{equation}
\label{eq:complete}
\int ^\infty _0 |k\rangle \langle k|k^{2}dk=1.
\end{equation}
We here make an approximation for Eq.~(\ref{eq:eigenvalue_eq_k})
in a numerical integral form by introducing the adequate integral mesh points
$k_{i}$ and $k_{j}$ and discretizing $k\to k_{j}$ and $k'\to k_{i}$ as
\begin{equation}
\label{eq:eigenequation_k}
\sum_{j}\langle \bar{k_{i}}|
H_{0}+V|\bar{k_{j}}\rangle \langle\bar{k_{j}}|\Psi _{n}\rangle
=E_{n}\langle \bar{k_{i}}|\Psi _{n}\rangle ,
\end{equation}
where $|\bar{k}_{i}\rangle$ and $|\bar{k}_{j}\rangle$
represent the plane-wave basis states in the matrices.
Those grids are characterized by the mesh points $k_{i}$ and $k_{j}$.
Thus, $|\bar{k_{i}}\rangle$ is defined as
\begin{equation}
\label{eq:bar_k}
|\bar{k_{i}}\rangle =k_{i}\sqrt{W_{i}}|k_{i}\rangle
\end{equation}
with the plane-wave states $|k_{i}\rangle$
and the weight factors for the numerical integral $\sqrt{W_{i}}$.
They are normalized as
\begin{equation}
\label{normalize_k}
\langle \bar{k_{i}}|\bar{k_{j}}\rangle =\delta _{i,j}.
\end{equation}

We note here that the eigenvectors $|\Psi _{n}\rangle$
in Eq.~(\ref{eq:eigenvalue_eq}) can be expressed as
$|\Psi _{n}\rangle =|\phi _{n}\rangle+\omega |\phi _{n}\rangle$
in terms of the operator $\omega$ and the P-space components
$|\phi _{n}\rangle=P|\Psi _{n}\rangle$.
Thus, the formal solution of $\omega$ is given by
$\omega =\sum_{n}|\Psi _{n}\rangle \langle \tilde{\phi}_{n}|$
with the biorthogonal state $\langle \tilde{\phi}_{n}|$ of $|\phi _{n}\rangle$.
In order to obtain the matrix elements of $\omega$,
we first solve Eq.~(\ref{eq:eigenequation_k}) by diagonalizing the matrix
elements, using the basis states $|\bar{k_{i}}\rangle$.
Then, the matrix elements $\omega$ for the basis states
$|\bar{k_{p}}\rangle$ in the $P$ space and $|\bar{k_{q}}\rangle$
in the Q space are obtained as
\begin{equation}
\label{eq:omega_qp}
\langle \bar{k_{q}}|\omega|\bar{k_{p}}\rangle =
\sum_{n=1}^{d}\langle \bar{k_{q}}|Q|\Psi _{n}\rangle
\langle \tilde{\phi}_{n}|P|\bar{k_{p}}\rangle,
\end{equation}
where $d$ is the number of the basis states (the integral points)
in the P space.
As shown in Eq.~(\ref{eq:PQ}) the P space (low-momentum space) and Q space
(high-momentum space) are defined with a cutoff momentum $\Lambda$ as
$0<k\leq \Lambda$ and $\Lambda<k<\infty$, respectively.
The bra states $\langle \tilde{\phi}_{n}|$ are obtained through the matrix
inversion as $[\langle \tilde{\phi}_{n}|\bar{k_{p}}\rangle ]
\equiv [ \langle \bar{k_{p'}}|\phi _{n}\rangle ]^{-1}$
and satisfy the relations
$\sum_{k_{p}}\langle \tilde{\phi} _{n}|\bar{k_{p}}\rangle
\langle \bar{k_{p}}|\phi _{n'}\rangle =\delta _{n,n '}$
and
$\sum_{n}\langle \bar{k_{p'}}|\tilde{\phi} _{n}\rangle
\langle \phi _{n}|\bar{k_{p}}\rangle =\delta _{k_{p'},k_{p}}$.
It should be noted that the solution $\omega$ given in Eq.~(\ref{eq:omega_qp})
is ambiguous in the sense that how to choose the set of eigenvectors
$\{ |\Psi _{n}\rangle , n=1, 2,..., d\}$ is not unique.
We here select $\{ |\Psi _{n}\rangle \}$ so that they have
the largest P-space overlaps $O_{n}
=\sum _{i=1}^{d}|\langle \Psi _{n}|P|\bar{k_{i}}\rangle|^{2}$ among
all the eigenstates in Eq.~(\ref{eq:eigenequation_k}).
As we will show later, the numerical calculation shows that this selection
of $\{ |\Psi _{n}\rangle \}$ leads to the same solution obtained from method-1.

In order to obtain the LMNN interaction (P-space effective interaction),
we introduce the eigenvalue equation for $\omega ^{\dagger}\omega$
in the P space as
\begin{equation}
\label{eq:omega2}
\omega ^{\dagger}\omega|\psi _{\alpha}\rangle
=\mu _{\alpha}^{2}|\psi _{\alpha} \rangle .
\end{equation}
Using the solutions to the above equation, the LMNN interaction of
a Hermitian type is given by~\cite{Suzuki82, Kuo93}
\begin{eqnarray}
\label{V_lowk}
\lefteqn{\langle \psi _{\alpha}|V_{{\rm low}-k}|\psi_{\beta}\rangle}
\nonumber \\
&=&\frac{\sqrt{(1+\mu _{\alpha}^{2})}
\langle \psi _{\alpha}|R|\psi _{\beta}\rangle
+\sqrt{(1+\mu _{\beta}^{2})}\langle \psi _{\alpha}|R^{\dagger}
|\psi _{\beta}\rangle}{\sqrt{(1+\mu _{\alpha}^{2})}
+\sqrt{(1+\mu _{\beta}^{2})}},\nonumber \\
\end{eqnarray}
where
\begin{equation}
\label{eq:V_nonhermitian}
R=P(V+V\omega)P
\end{equation}
is a low-momentum (effective) interaction of a non-Hermitian type.
Finally, the matrix elements of the LMNN interaction using the plane-wave
basis states $|k_{i}\rangle$ and $|k_{j}\rangle$ are obtained as
\begin{equation}
\label{ME_V_lowk}
\langle k_{i}|V_{{\rm low}-k}|k_{j}\rangle=\frac{\sum_{\alpha,\beta}
\langle \bar{k}_{i}|\psi _{\alpha}\rangle\langle \psi _{\alpha}|V_{{\rm low}-k}
|\psi _{\beta}\rangle \langle \psi _{\beta}|
\bar{k}_{j}\rangle}{k_{i}k_{j}\sqrt{W_{i}W_{j}}}.
\end{equation}
By an interpolation technique the elements of the potential are
prepared at arbitrary momenta in the P space.

\section{\label{sec:results}Results and discussion}

As mentioned in the previous section, we have two different methods
based on the unitary transformation. Here we  make the cross
check using both methods.
\begin{figure}[tt]
\includegraphics[width=.300\textheight]{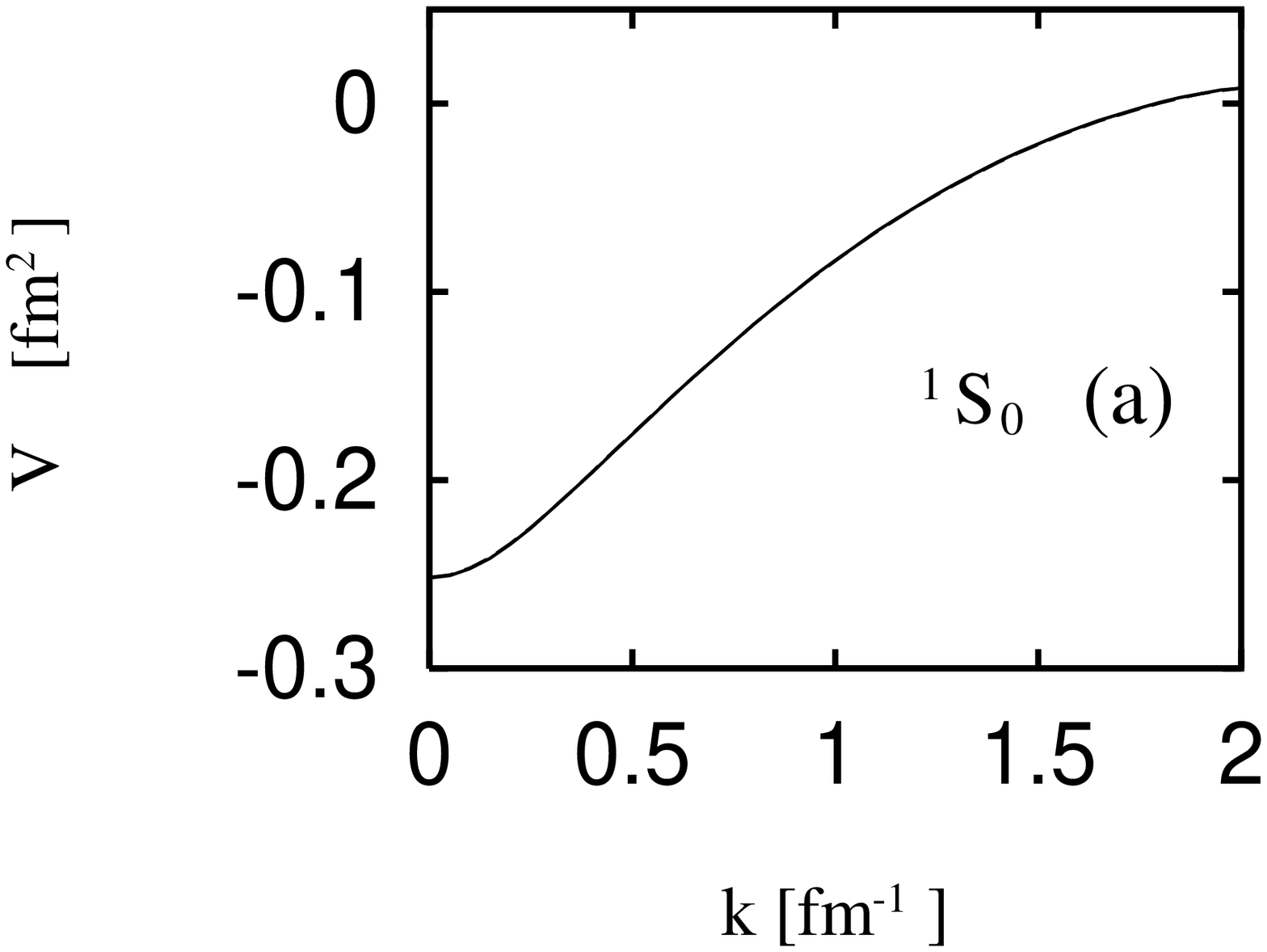}
\includegraphics[width=.300\textheight]{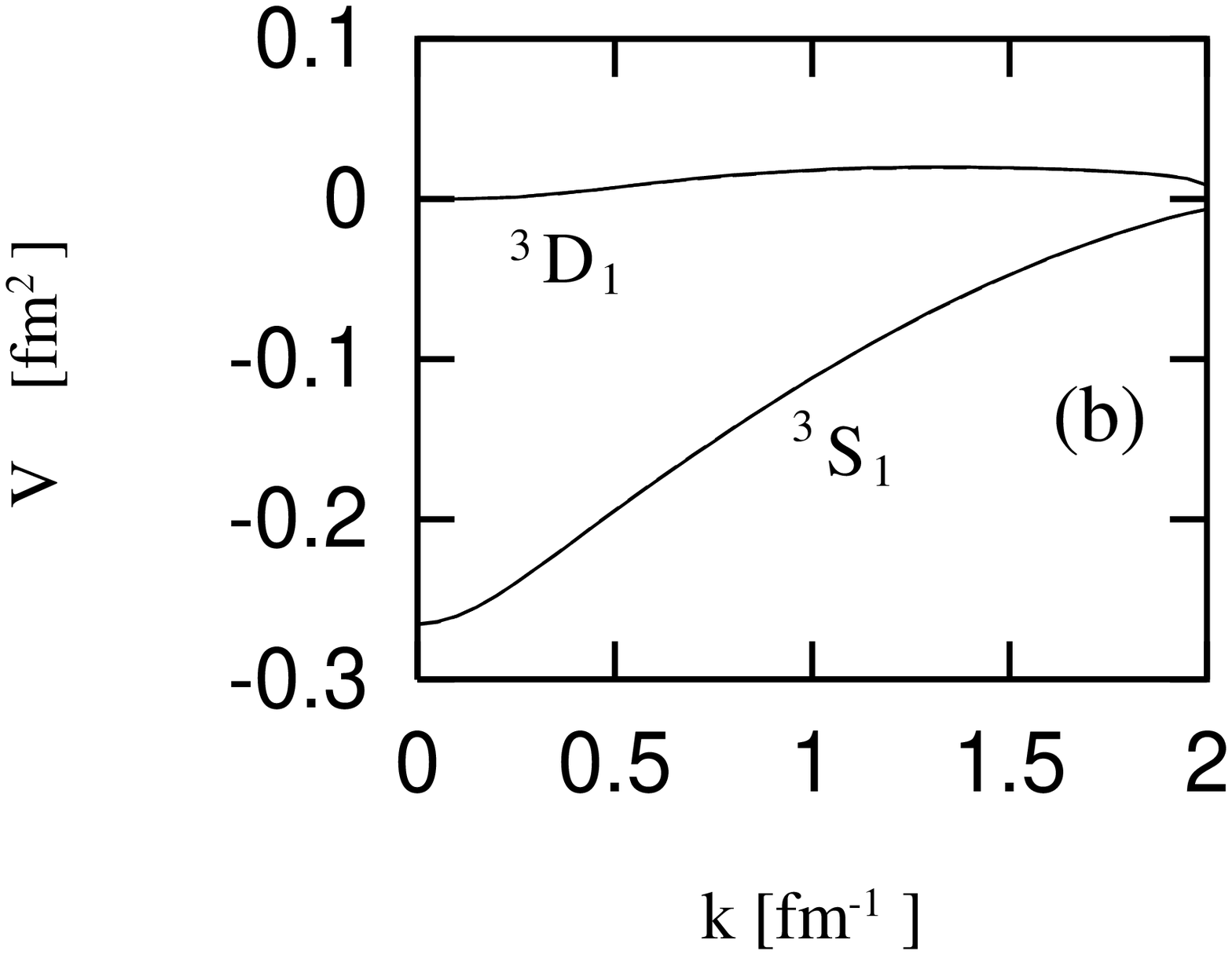}
\caption{\label{fig:ME} Comparison of LMNN interactions from the CD-Bonn
potential in the case of $\Lambda =2.0$ fm$^{-1}$.
The diagonal matrix elements $\langle
k|V_{{\rm low}-k}|k\rangle$ for the $^{1}S_{0}$ (a) and
$^{3}S_{1}$-$^{3}D_{1}$ (b) partial waves are shown.
The lines depict the method-1 (solid) and the method-2
(long-dashed) results.
Because they coincide very well,
one cannot distinguish both lines by the eye.}
\end{figure}

\begin{figure}[tt]
\includegraphics[width=.300\textheight]{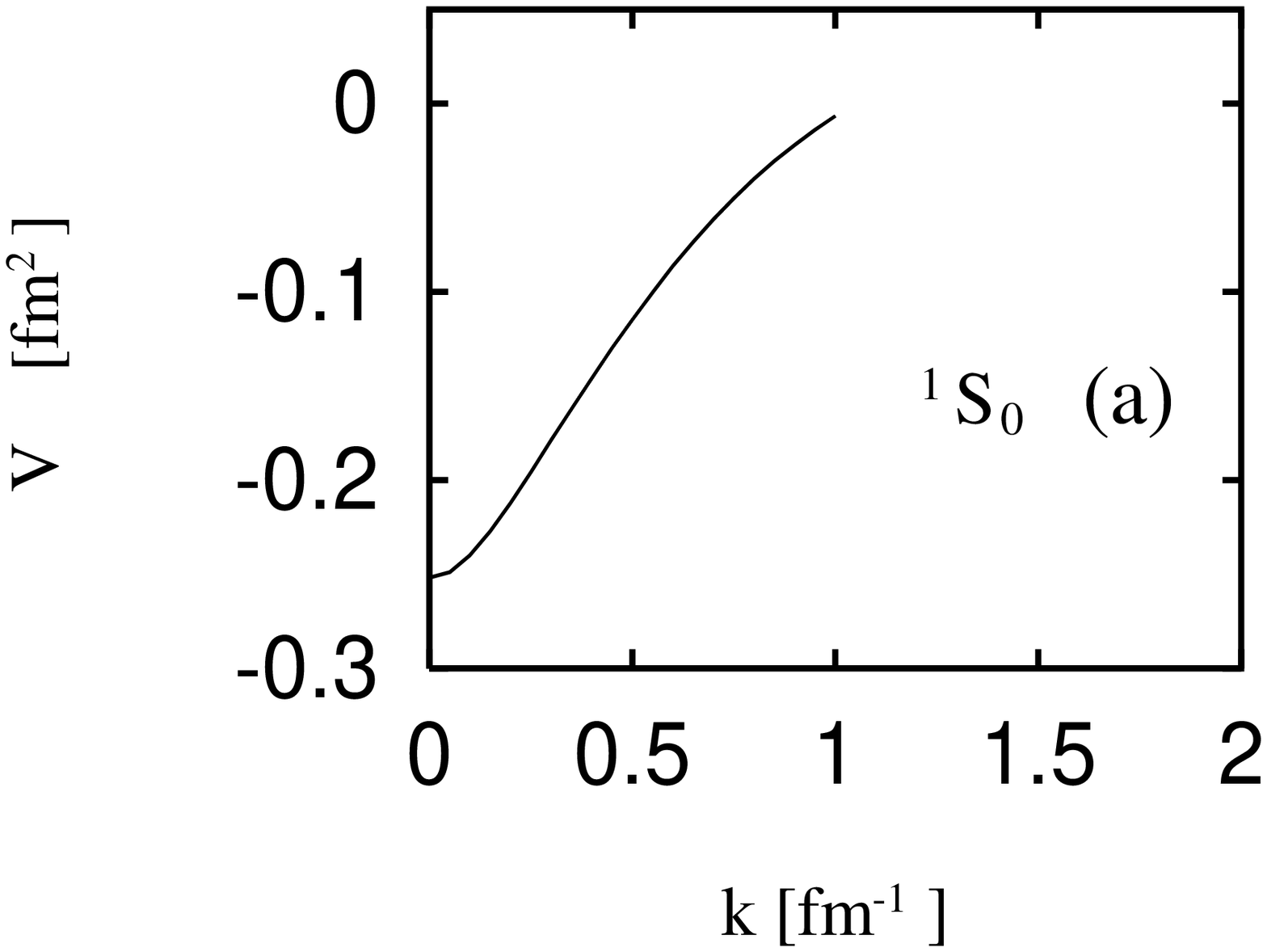}
\includegraphics[width=.300\textheight]{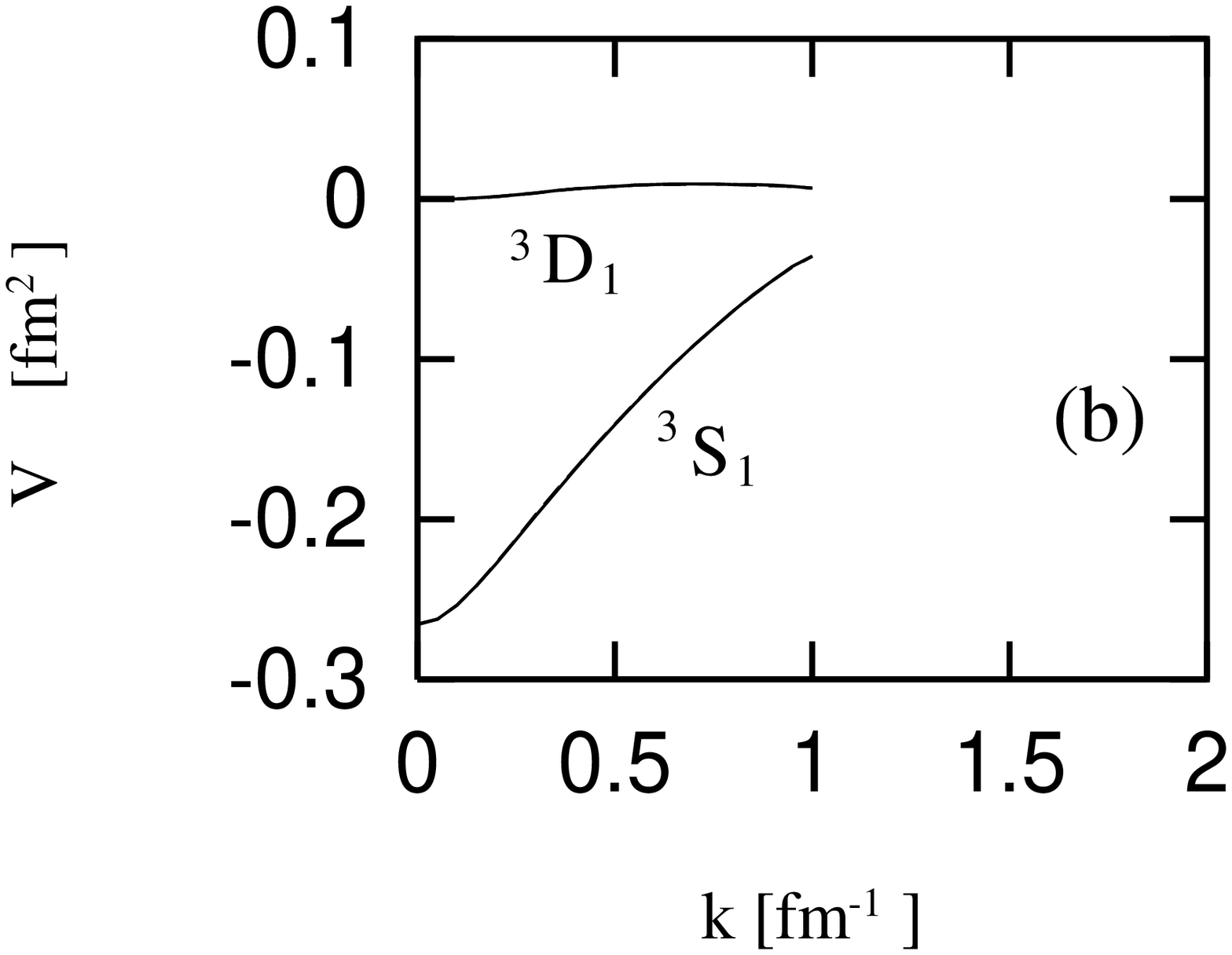}
\caption{\label{fig:ME1} Comparison of LMNN interactions for
the off-diagonal elements  $\langle k|V_{{\rm low}-k}| 2k\rangle$.
Description is the same as in Fig.~\ref{fig:ME}.}
\end{figure}

\begin{figure}[tt]
\includegraphics[width=.300\textheight]{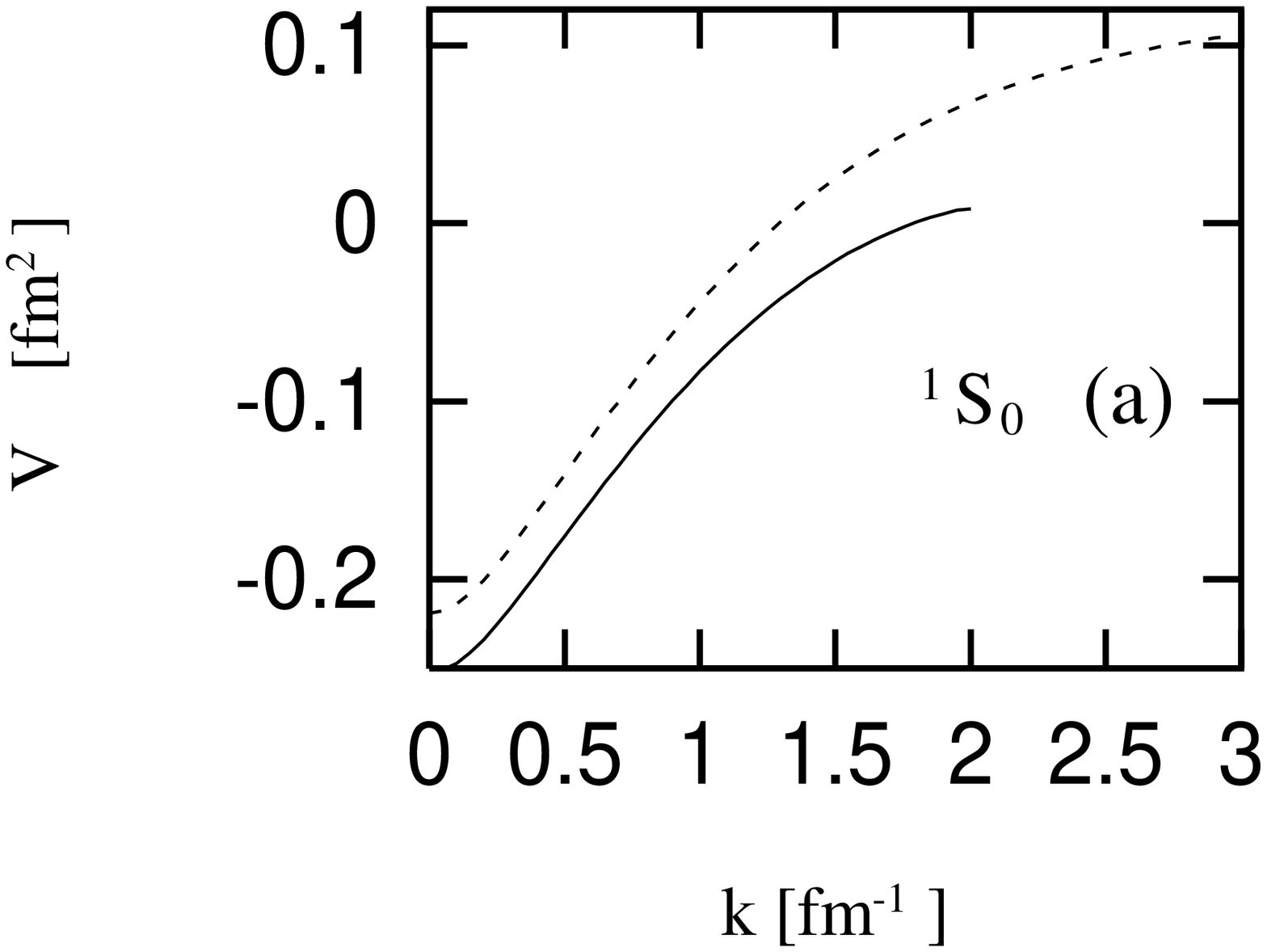}
\includegraphics[width=.300\textheight]{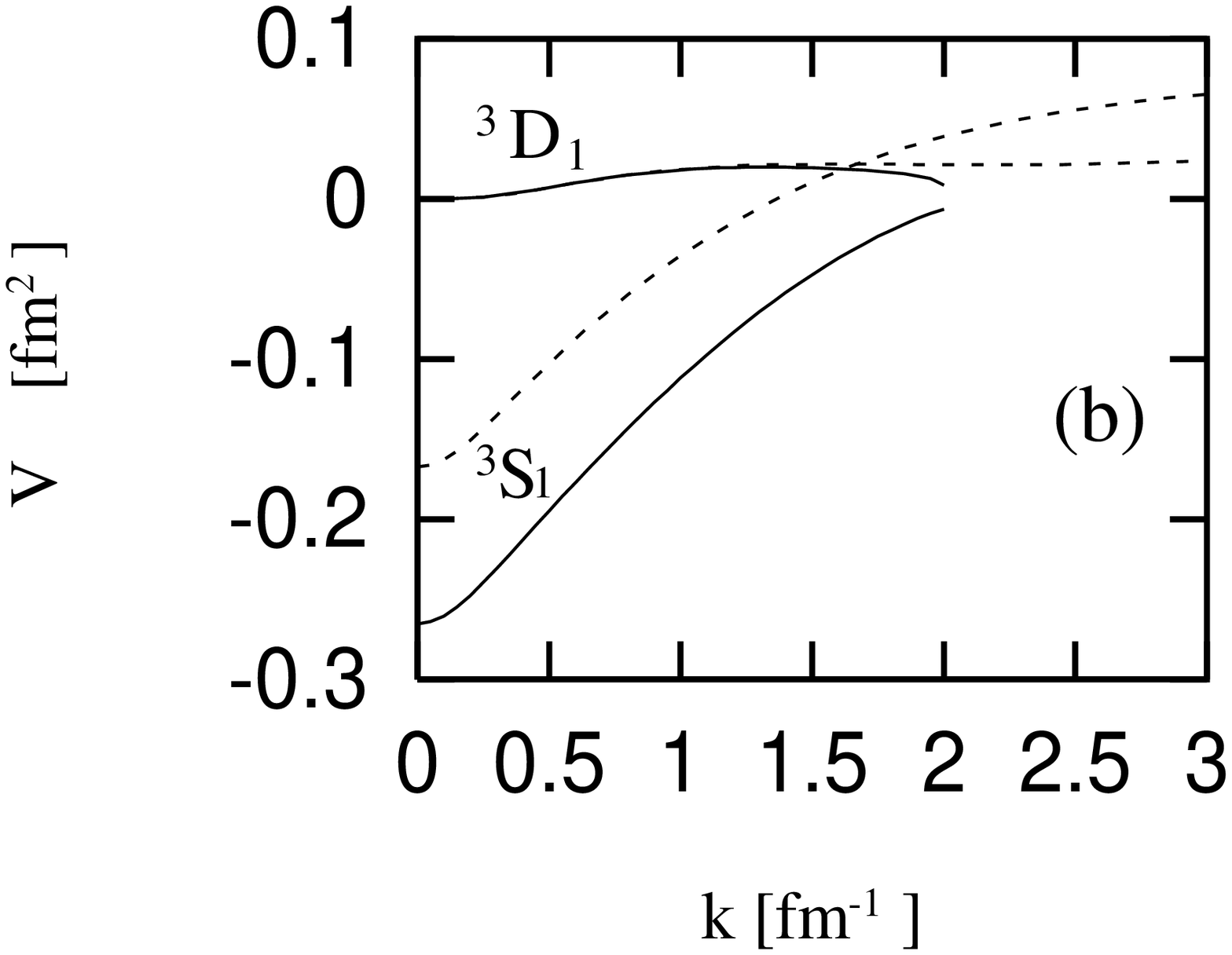}
\caption{\label{fig:ME2} Comparison of the LMNN interaction
in the case of $\Lambda= 2.0$ fm$^{-1}$
and the original CD-Bonn potential for the $^{1}S_{0}$ (a)
and $^{3}S_{1}$-$^{3}D_{1}$ (b) partial waves.
The solid and dashed lines depict the diagonal matrix elements of
the LMNN interaction and the original CD-Bonn potential, respectively.}
\end{figure}

In Fig.~\ref{fig:ME}, the diagonal matrix elements of the LMNN interactions
for the neutron-proton $^{1}S_{0}$ and $^{3}S_{1}$-$^{3}D_{1}$ channels using
the CD-Bonn potential~\cite{Machleidt96} are shown in the case of
$\Lambda =2.0$ fm$^{-1}$.
In order to see the off-diagonal matrix elements, we also illustrate
$\langle k|V_{{\rm low}-k}|2k\rangle$ in Fig.~\ref{fig:ME1}.
One can see that the results obtained by the two methods are almost
the same for both the diagonal and non-diagonal matrix elements within $3$-$4$
digits using typically $100$ integral grid points.
Having obtained the same results with high precision using two very different
methods, we have confidence in our numerical results.

\begin{figure}[t]
\includegraphics[width=.300\textheight]{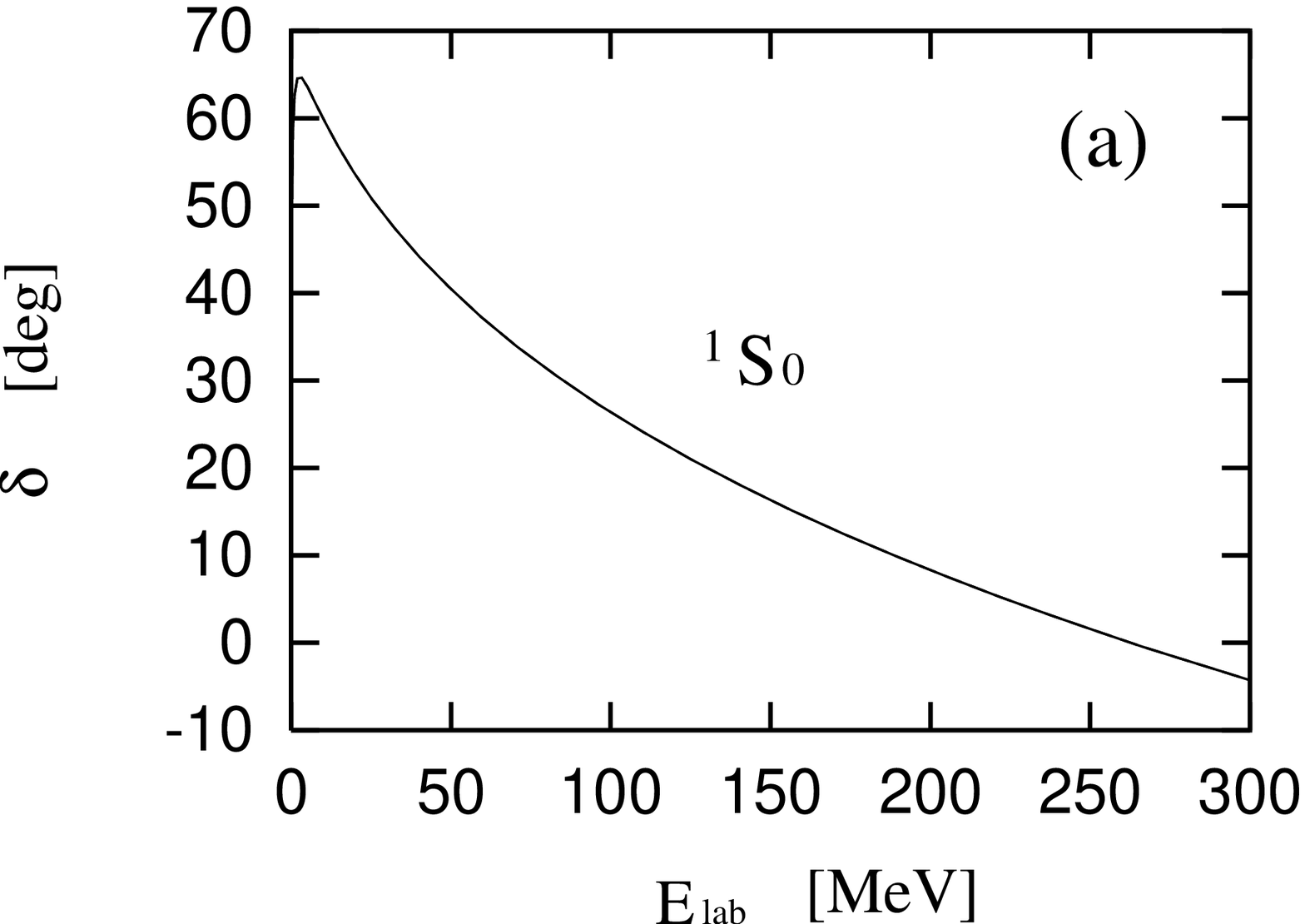}
\includegraphics[width=.300\textheight]{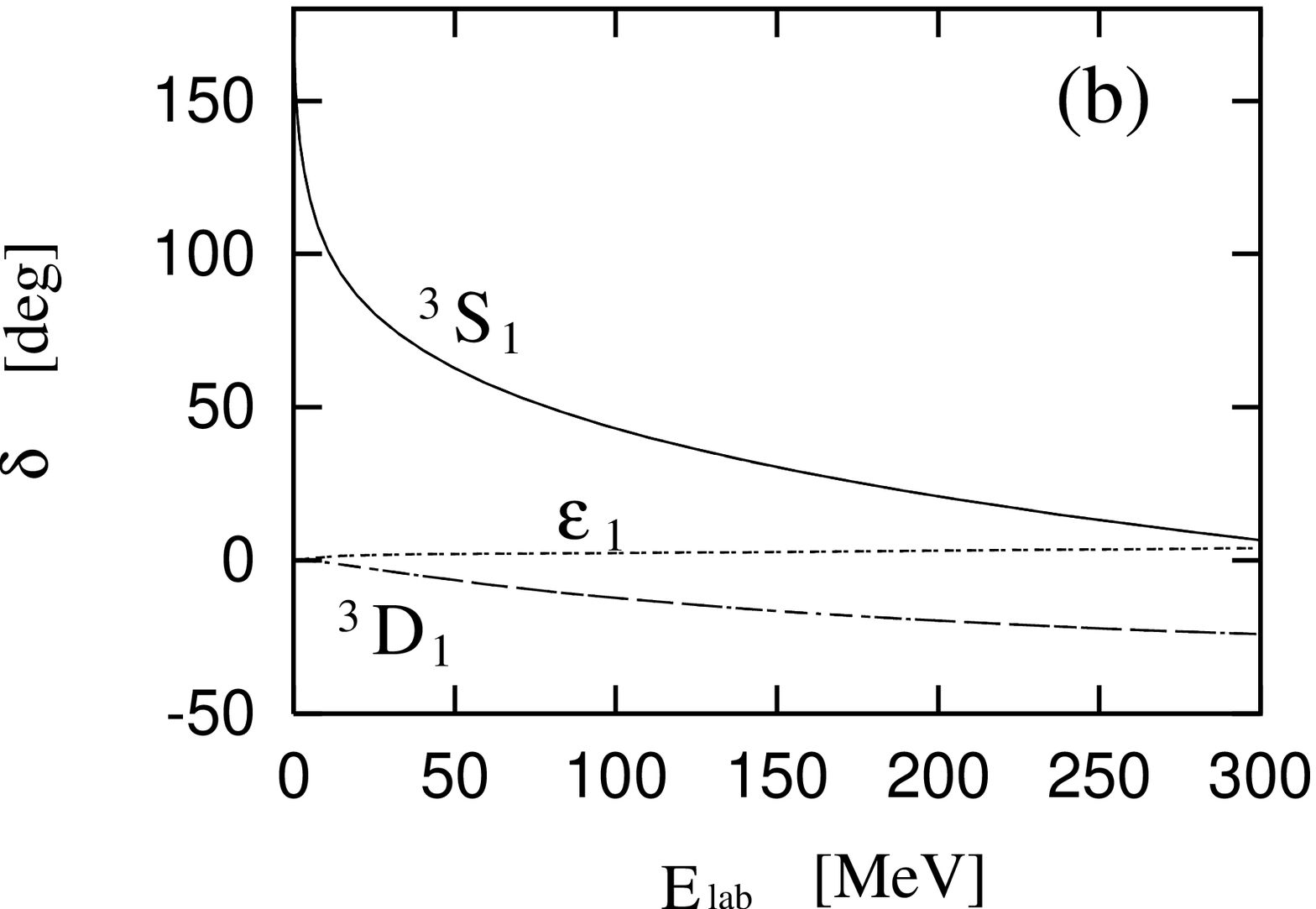}
\caption{\label{fig:phase} Phase shifts for the $^{1}S_{0}$ (a) and the
  $^{3}S_{1}$-$^{3}D_{1}$ (b)
channels below $E_{\rm lab}=300$ MeV.  
Because the lines from the LMNN interactions and the original CD-Bonn
potential coincide, one cannot distinguish both lines by the eye. 
In Fig.~(b) the upper, middle, and lower lines depict the $^{3}S_{1}$
phase shift, the mixing parameter $\epsilon_{1}$, and the $^{3}D_{1}$
phase shift, respectively. }
\end{figure}

In Fig.~\ref{fig:ME2}, we show the same matrix elements of the original
CD-Bonn potential and the LMNN interaction for the sake of comparison.
The LMNN potential is very different from the original one.
Nevertheless, the scattering phase shifts and the mixing parameter below
$E_{\rm lab}=300$ MeV for the $^{1}S_{0}$ and $^{3}S_{1}$-$^{3}D_{1}$ channels,
as shown in Fig.~\ref{fig:phase},
reproduce exactly the ones obtained from the original interaction.
This has been also shown by Bogner {\it et al.}~\cite{Bogner03}.

We now regard deuteron properties.
In the light of the unitary transformation for the two-nucleon system,
all the calculated deuteron binding energies for various values of $\Lambda$
must reproduce the exact value using the original interaction.
In Table~\ref{tab:deuteron}, calculated deuteron binding energies for various
$\Lambda$ using the CD-Bonn~\cite{Machleidt96} and the Nijm-I~\cite{Stoks94}
potentials are tabulated together with deuteron $D$-state probabilities.
Note that the $D$-state probability is not observable~\cite{Friar79}.
The values in the last row are the exact values quoted from the original
papers of the CD-Bonn and the Nijm-I potentials.
Indeed, we see that the calculated binding energy for each $\Lambda$ reproduces
the exact value with high accuracy.
However, as for the $D$-state probability, the difference between the results
using the LMNN interaction and the exact value becomes larger as the value of
$\Lambda$ becomes smaller.
If one uses the corresponding effective operator
in calculating the $D$-state probability,
i.e. the unitarily transformed projection operator onto the $D$ state,
then one would reproduce the original value for this quantity.

\begin{table}[t]
\caption{\label{tab:deuteron} Calculated binding energies $E_{\rm b}$(MeV) and
$D$-state probabilities $P_{D}(\%)$ of the deuteron for various values of
$\Lambda$.
The values in the last row are those quoted from the original
papers of the CD-Bonn and
the Nijm-I potentials.}
\begin{ruledtabular}
    \begin{tabular}{ccccc}
 & \multicolumn{2}{c}{CD Bonn} & \multicolumn{2}{c}{Nijm I}\\
 $\Lambda$(fm$^{-1}$) & $E_{\rm b}$(MeV) & $P_{D}(\%)$ & $E_{\rm b}$(MeV) &
$P_{D}(\%)$ \\
\hline
 $1.0$ & $-2.224576$ & $1.21$ & $-2.224575$ & $1.24$ \\
 $2.0$ & $-2.224576$ & $3.55$ & $-2.224575$ & $3.83$ \\
 $3.0$ & $-2.224576$ & $4.55$ & $-2.224575$ & $5.12$ \\
 $4.0$ & $-2.224576$ & $4.79$ & $-2.224575$ & $5.53$ \\
 $5.0$ & $-2.224576$ & $4.83$ & $-2.224575$ & $5.64$ \\
 $6.0$ & $-2.224576$ & $4.83$ & $-2.224575$ & $5.66$ \\
 $7.0$ & $-2.224576$ & $4.83$ & $-2.224575$ & $5.66$ \\
quoted & $-2.224575$ & $4.83$ & $-2.224575$ & $5.66$ \\
    \end{tabular}
\end{ruledtabular}
\end{table}

As shown in Table~\ref{tab:deuteron}, for the deuteron, the difference of
the wave functions does not affect the binding energy since we perform
the unitary transformation for the two-nucleon system.
However, if we apply the LMNN interaction to the calculation of
the ground-state energies of many-body systems, the situation will change.
This is because the unitary transformation in the two-nucleon Hilbert space is
not unitary any more in the Hilbert space of three and more nucleons.
As a consequence, calculated binding energies will depend on $\Lambda$.
In order to examine the $\Lambda$ dependence,
we perform the Faddeev and Yakubovsky calculations for few-nucleon systems.
Recent precise calculations for few-nucleon systems are reviewed
in Refs.~\cite{Kamada01,nogga03a}.

\begin{figure}[t]
\includegraphics[width=.360\textheight]{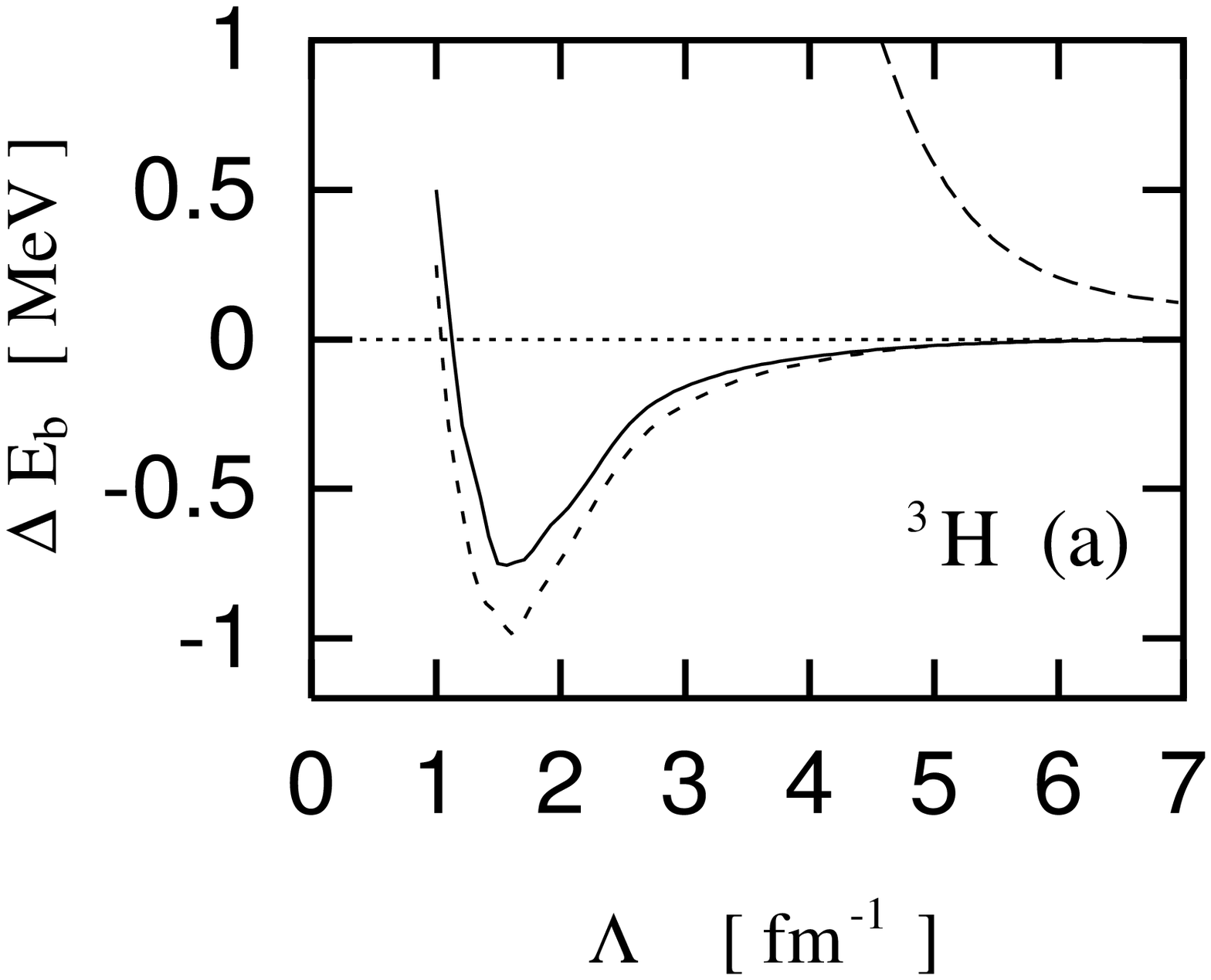}
\includegraphics[width=.360\textheight]{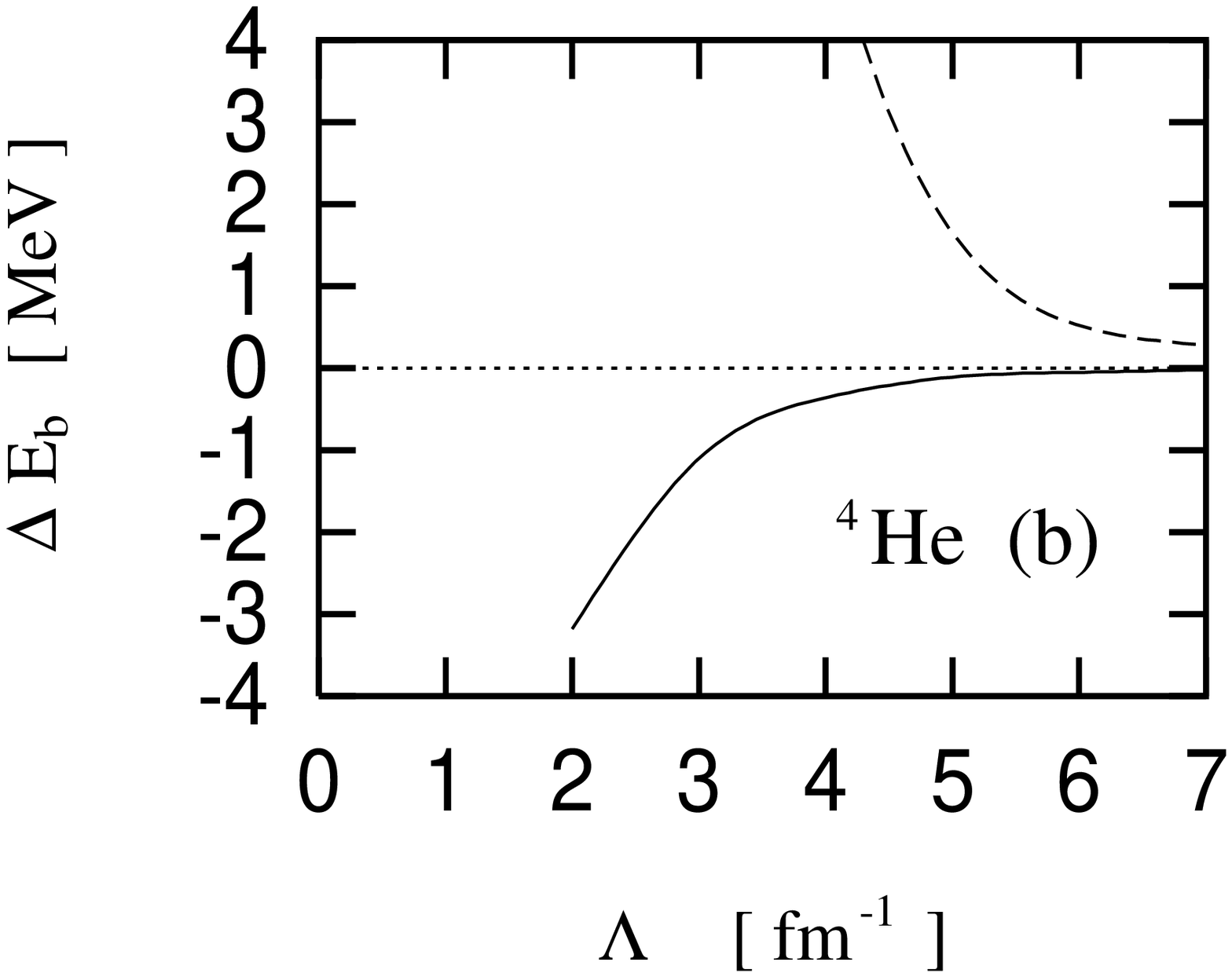}
\caption{\label{fig:H3_He4} Energy shifts $\Delta E_{\rm b}$ of
$^{3}$H (a) and $^{4}$He (b) as a function of $\Lambda$
using the LMNN interaction (solid) from the CD-Bonn potential.
The long-dashed 
lines in (a) and (b) are plotted for the case of simple momentum
cutoff calculations.
The short-dashed line in (a) depicts the result for the LMNN
interaction from the Nijm-I potential.
}

\end{figure}

Figure~\ref{fig:H3_He4}(a) exhibits the energy shift
$\Delta E_{\rm b}\equiv E_{\rm b}(\Lambda)-E_{\rm b}(\infty)$
from the ground-state energy $E_{\rm b}(\infty)$ of $^{3}$H as a function of
$\Lambda$ based on the CD-Bonn potential (solid) and the Nijm-I one
(short-dashed) by a 34-channel Faddeev calculation.
In the present study, only the neutron-proton interaction is used for all the
channels for simplicity.
The exact value $E_{\rm b}(\infty)$
using the original potential CD Bonn (Nijm I) is $-8.25$ ($-8.01$) MeV.
The long-dashed line depicts the Faddeev calculation
using the original interaction, where the high-momentum components
beyond $\Lambda$ are simply truncated.
The numerical stability is lost within the area $\Lambda \le 1$ fm$^{-1}$.
For the case of the CD-Bonn potential, we need $\Lambda > 8$ fm$^{-1}$ to reach
the exact value if the accuracy of $100$ keV is required
for the case of simple truncation.
This situation is largely improved if we use the LMNN interaction.
Even if we require the accuracy of $1$ keV, we do not need the high-momentum
components beyond $\Lambda \sim 8$ fm$^{-1}$.
However, it should be noted that the results using the LMNN interaction for
the values smaller than $\Lambda \sim 5$ fm$^{-1}$ vary considerably,
and there occurs the energy minimum around $\Lambda \sim 2$ fm$^{-1}$.
The cutoff value, which produces the minimal value of $\Delta E_{\rm b}$,
is close to the value proposed by Bogner {\it et al.}~\cite{Bogner03}.

We note here that the wave function using the LMNN interactions
is very close to the true wave function if the Jacobi momentum set
$(p,q)$ of the three-nucleon system is smaller than the adopted value of
$\Lambda$ when we take $\Lambda \ge 4$ fm$^{-1}$.

By solving the Yakubovsky equations
we can calculate the binding energy of $^{4}$He.
A similar tendency of the $\Lambda$ dependence for $^{3}$H can be
also seen in the results for $^{4}$He.
In Fig.~\ref{fig:H3_He4}(b), the $\Lambda$ dependence of the energy
shift $\Delta E_{\rm b}$ of the ground state of $^{4}$He is illustrated.
We demonstrate only the case of the CD-Bonn potential.
Here we adopt the $S$-wave ($5+5$-channel) approximation and do not
include the Coulomb force for simplicity~\cite{Kamada01}.
We also use only the neutron-proton interaction for all the channels.
The exact value $E_{\rm b} (\infty)$ using the original CD-Bonn potential
in the above-mentioned approximations is $-27.74$ MeV.
The shape of the curve is similar to that for $^{3}$H up to
$\Lambda \sim 2$ fm$^{-1}$.
The numerical instability already starts at $\Lambda \leq 2$ fm$^{-1}$.
Here it seems to conserve the strong correlation
relation~\cite{Kamada01}
between $\Delta E_{\rm b}(^{4}{\rm He})$ and $\Delta E_{\rm b}
(^{3}\rm {H})$ where the ratio is about 5 in the region $\Lambda \ge
3$ fm$^{-1}$.  

In the case of $^{3}$H, the energy shift $\Delta E_{\rm b}$ is about $750$
keV ($1$ MeV in case of Nijm-I potential) at the minimum point.
On the other hand, the difference amounts to about $3$ MeV for $^{4}$He at
$\Lambda = 2.0$ fm$^{-1}$.
We remark that the minimal value of the cutoff $\Lambda$ which leads only to
small deviations of the $^{3}$H and $^{4}$He binding energies from their exact
values is $\Lambda \sim 5$ fm$^{-1}$.
As far as the excitation spectra of low-lying states from the ground 
state are concerned, 
the $\Lambda$ dependence at $\Lambda \sim 2$ fm$^{-1}$ may be weak
and the LMNN interactions could be useful as has been shown
in the shell-model calculations~\cite{Bogner02}.

It is noted that 
for LMNN interactions considered as a purely computational tool,
which allows to get rid of the hard core and thus enables many-body
calculations based on realistic 
NN potentials without using the $G$-matrix formalism, we recommend the
value of the cutoff $\Lambda$ 
of the order of (or higher than) $\sim 5 $ fm$^{-1}$. One then
ensures that the binding energies do not deviate significantly from 
their exact values  in the case of few-nucleon systems.
Further reducing the values of the cutoff $\Lambda$ leads to a
significant suppression of the repulsive part of 
the interaction which becomes visible.
In another words,
if one would be able to extend the unitary transformation formalism to
more-than-two-nucleon Hilbert space, additional three- and 
more-nucleon forces would be generated through eliminating the
high-momentum components. These forces would 
probably be repulsive~\footnote{Speaking more precisely, they would
have an affect of decreasing the binding energies.} 
(at least for the systems discussed in this work) and would restore the values
of the few- and many-nucleon binding energies and other observables
found in calculations with the 
original, not transformed NN forces.

On the other hand, it is well known that
all realistic NN interactions underbind light nuclei such as $^3$H and
$^4$He. As a consequence, attractive three-nucleon forces 
are needed in order to reproduce the experimental numbers. It is,
therefore, in principle possible, that the repulsive 
effective many-nucleon forces generated through elimination of the
high-momentum components compensate, to some 
extent, these missing attractive forces, minimizing the total effect
of the many-nucleon interactions. 
If this oversimplified picture reflects the real situation, one would
observe a better agreement with the data for calculations based on the
LMNN interactions as compared to the ones based on 
realistic NN forces. Although the results provided by existing many-body 
calculations as well as by the present few-body
calculations with LMNN interactions appear to be
quite promising in this respect~\footnote{Notice however the counter
examples mentioned before.}, 
and thus might speak in favor of the above-mentioned tendency of
counteracting these two 
kinds of many-body forces, to the best of our knowledge,
no general proof of the validity of the above-mentioned picture has
yet been offered. 

In addition, one should
keep in mind that only a restricted information about many-nucleon
forces, which might have very complicated spin and spatial 
structure, is provided by the discrete spectrum. A much better testing
ground is 
served by a variety of scattering observables.
It would therefore be interesting to test the LMNN forces in the three-nucleon
continuum. Further, we notice that the choice of the LMNN 
is not unique. Preserving the half-on-shell $T$ matrix does not seem
to be advantageous in any respect, at least from the conceptual 
point of view. If this condition is not required, infinitely many
equivalent LMNN can be constructed by means of a 
unitary transformation in the low-momentum subspace. Even if many-body
effects appear to be small for one particular choice 
of LMNN forces, this might not be the case for a different choice.

Certainly, the LMNN interactions with small values of the cutoff, $\Lambda
\sim 2 $ fm$^{-1}$, are not disadvantageous compared to existing
realistic NN forces 
regarding the nuclear structure calculations,
if these realistic forces are viewed as purely phenomenological
parametrizations with no physical content, 
underlying the only requirement of reproducing properly the low-energy NN data.

\section{\label{sec:summary}Summary}

The LMNN interactions have been derived through a unitary-transformation theory
from realistic nucleon-nucleon interactions
such as the CD-Bonn and the Nijm-I potentials.
We have constructed the LMNN interactions by two different methods
which are based on the common unitary transformation.
In order to have a cross check of both computer codes,
we have shown that the LMNN interactions obtained by the two methods
yield the same results.
The LMNN interaction reproduces the low-energy observables in the two-nucleon
system with high precision, which has been confirmed
by the calculation of the deuteron binding energy and the phase shifts.

The LMNN interaction has been successfully applied to the Faddeev-Yakubovsky
calculations for three- and four-nucleon systems.
The calculated binding energies of the few-nucleon systems begin to deviate
from the values calculated using the original NN potentials for $\Lambda$
smaller than $\sim 5$ fm$^{-1}$, whereas the results obtained by simply cutting
off the high momentum components without performing a unitary transformation
deviate considerably even at much higher values of $\Lambda$.
In an appropriately truncated (i.e. with $\Lambda \geq 5$ fm$^{-1}$)
low-momentum space the LMNN interaction reproduces the exact values of
the binding energies, at least, for the few-nucleon systems.

However, we should keep in mind that the calculations of ground-state energies
using the LMNN interaction for $\Lambda \sim 2$ fm$^{-1}$ yield
considerably more attractive results than the exact values.
We note that as shown in Fig. 1 of Kuckei {\it et al.}'s work~\cite{Kuckei}
one needs more than $4.0$ fm$^{-1}$ as the cutoff value $\Lambda$ 
in order to reproduce at least qualitatively the saturation property
of nuclear matter.
Thus, the application of the LMNN interaction to structure calculations
should be done with care, though the LMNN interaction for
$\Lambda \sim 2$ fm$^{-1}$ may be suitable for the calculation of
the excitation spectra of low-lying states as has been shown
in the shell-model calculations~\cite{Bogner02}.

\begin{acknowledgments}

One of the authors (S.~F.) acknowledges the Special Postdoctoral Researchers
Program of RIKEN.
This work was supported by a Grant-in-Aid for Scientific Research (C)
(Grant No. 15540280) from Japan Society for the Promotion of Science,
a Grant-in-Aid for Specially Promoted Research (Grant No. 13002001)
from the Ministry of Education, Culture, Sports, Science and Technology
in Japan,  and the
U.S. Department of Energy Contract No. DE-AC05-84ER40150 under which
the Southeastern Universities Research Association (SURA) operates the
Thomas Jefferson Accelerator Facility.
The numerical calculations were performed on Hitachi SR8000
(Leibnitz-Rechenzentrum f\"ur die M\"unchener Hochschule) in Germany.

\end{acknowledgments}

\end{document}